\documentclass[amsmath,amssymb,amssymb]{revtex4-1}

\usepackage{graphicx,epsfig,epstopdf,amssymb,amsmath,bbm}
\usepackage{mathrsfs}
\usepackage{dcolumn}
\usepackage{bm}
\usepackage{natbib}
\begin{document}
\title{Linear and nonlinear frequency- and time-domain spectroscopy with multiple frequency combs}
\author{Kochise Bennett}
\email{kcbennet@uci.edu}
\author{Jeremy R. Rouxel}
\email{jrouxel@uci.edu}
\author{Shaul Mukamel}
\email{smukamel@uci.edu}
\affiliation{Chemistry Department and Physics and Astronomy Department, University of California, Irvine, California 92697-2025, USA}
\date{\today}
\begin{abstract}
Two techniques that employ equally spaced trains of optical pulses to map an optical high frequency into a low frequency modulation of the signal that can be detected in real time are compared.
The development of phase-stable optical frequency combs has opened up new avenues to metrology and spectroscopy. 
The ability to generate a series of frequency spikes with precisely controlled separation permits a fast, highly accurate sampling of the material response. 
Recently, pairs of frequency combs with slightly different repetition rates have been utilized to down-convert material susceptibilities from the optical to microwave regime where they can be recorded in real time.  We show how this one-dimensional dual comb technique can be extended to multiple dimensions by using several combs.  
We demonstrate how nonlinear susceptibilities can be quickly acquired using this technique. 
In a second class of techniques, sequences of ultrafast mode locked laser pulses are used to recover pathways of interactions contributing to nonlinear susceptibilities by using a photo-acoustic modulation varying along the sequences. We show that these techniques can be viewed as a time-domain analogue of the multiple frequency comb scheme.
\end{abstract}
\maketitle

\section{Introduction}

Mode-locked lasers, widely used for generating ultrashort light pulses,
operate by keeping a fixed phase relationship between the modes of the laser, leading to periodic constructive interferences that result in a train of pulses with a well-defined interpulse separation $T$. 
In the frequency domain, such an electric field corresponds to a series of frequency spikes with fixed wavenumber separation $1/T$, called a frequency comb \cite{udem1999accurate,holzwarth2000optical,ye2005femtosecond,cundiff2002phase,cundiff2003colloquium}.  Their advent has revolutionized high-precision metrology \cite{udem2002optical,cundiff2003colloquium}, enabled the creation of intense, few-cycle pulses with controlled carrier-envelope phase \cite{baltuvska2003attosecond,krausz2009attosecond}, and shows great promise for novel spectroscopic applications \cite{holzwarth2000optical,cundiff2003colloquium,coddington2008coherent,bernhardt2010cavity,glenn2014nonlinear, ideguchi2014adaptive,coddington2010coherent}. 
Frequency combs can be used for fast data acquisition and also offer ultrahigh resolution spectra.  

In a standard ultrafast nonlinear spectroscopy set-up, multiple pulses generate several interaction pathways within the matter that may then be separated through various techniques (phase-cycling, phase matching). The laser's repetition rate is then used to accumulate signals with a satisfactory signal to noise ratio. In the scheme presented here, all the pulses in the comb sequence are used to intrinsically isolate a desired contribution to the signal. A variety of spectroscopic signals have been measured using dual frequency combs (DFC).
Traditionally, experiments have been conducted by mixing the two lasers, with either one or both having travelled through the material, and measuring the interference between the two \cite{cundiff2003colloquium,schliesser2012mid,diddams2007molecular}.  The Fourier transform of this interferogram gives the signal that, at linear order, is the absorption spectrum and higher-order contributions showing Raman and other resonances \cite{ideguchi2012raman,ideguchi2013coherent}. 
The use of multiple frequency combs with slightly different frequency spacings induces slow temporal modulations of the various pathways contributing to the signal\cite{lomsadze2017four}. 
These down-shifted frequencies can be small enough to allow their detection by electronic means in real time while conserving the high frequency resolution of the comb. 

This technique can be alternatively described in the time domain. 
Since a frequency shift is equivalent to a modulation in the time domain, the advantages of using a frequency comb can also be, in principle, obtained by modulating pulses within a sequence with a varying frequency\cite{Bruder:17,PhysRevA.92.053412}. This technique has been used\cite{doi:10.1063/1.2386159,doi:10.1063/1.2800560} to detect and select various interaction pathways in two-photon fluorescence. 
The modulation of the pulses is achieved by using an independent acousto-optic modulator (AOM) for each of the four pulses in the interaction scheme. A standard ultrafast laser can be used in this application and the modulation of the pulses is tailored to match the repetition rate of the laser, releasing the constraint of using a comb in the frequency domain. 
On the other side, frequency combs offer a frequency shift control that would be difficult to achieve by modulating standard sources in the time domain.

In this manuscript, we develop a unified description for the multi-comb spectroscopies and the phase-modulated AOM detection techniques and compare the respective signals.
Both techniques have their merits and limitations. AOM is more straightforward to implement since it requires a simpler ultrafast laser\cite{nardin2013multidimensional}: a typical laser oscillator at 100MHz can be used and the downshifted signals are acquired at 3 to 13 kHz using a lock-in amplifier. However, phase modulation using AOM still requires varying the delays between the pulses. This requires long acquisition times because of the necessity to scan various delay stages as well as requiring a larger setup.
Frequency comb techniques, on the other hand, do not require scanning any delays and the acquisition can be done at the much higher frequency of the comb repetition rate.
The downshifting of the relevant signals is due to the frequency shift between various combs. The signal can then be modulated at very low, few-Hertz frequencies.

We extend the formalism of multidimensional spectroscopy to account for incoming fields composed of multiple combs. This can be used to analyze the aforementioned experiments and to guide the development of these techniques.
In section \ref{sec:signals}, we derive general expressions for nonlinear signals using broadband transmission and fluorescence detection.  Our derivations make clear that the same ideas of temporal signal modulation can be applied to incoherent detection when the exciting field is an overlap of combs.  
In section \ref{sec:expand}, we perturbatively expand the signal to linear and third order to demonstrate the effect of using a DFC field. Four-comb spectroscopic techniques for measuring the third order response function are discussed in section \ref{sec:4comb}.
Finally, in section \ref{section4}, we compare the phase modulation of ultrafast pulses with the frequency comb techniques.

\begin{figure}[h!]
\centering
\includegraphics[width=0.8\textwidth]{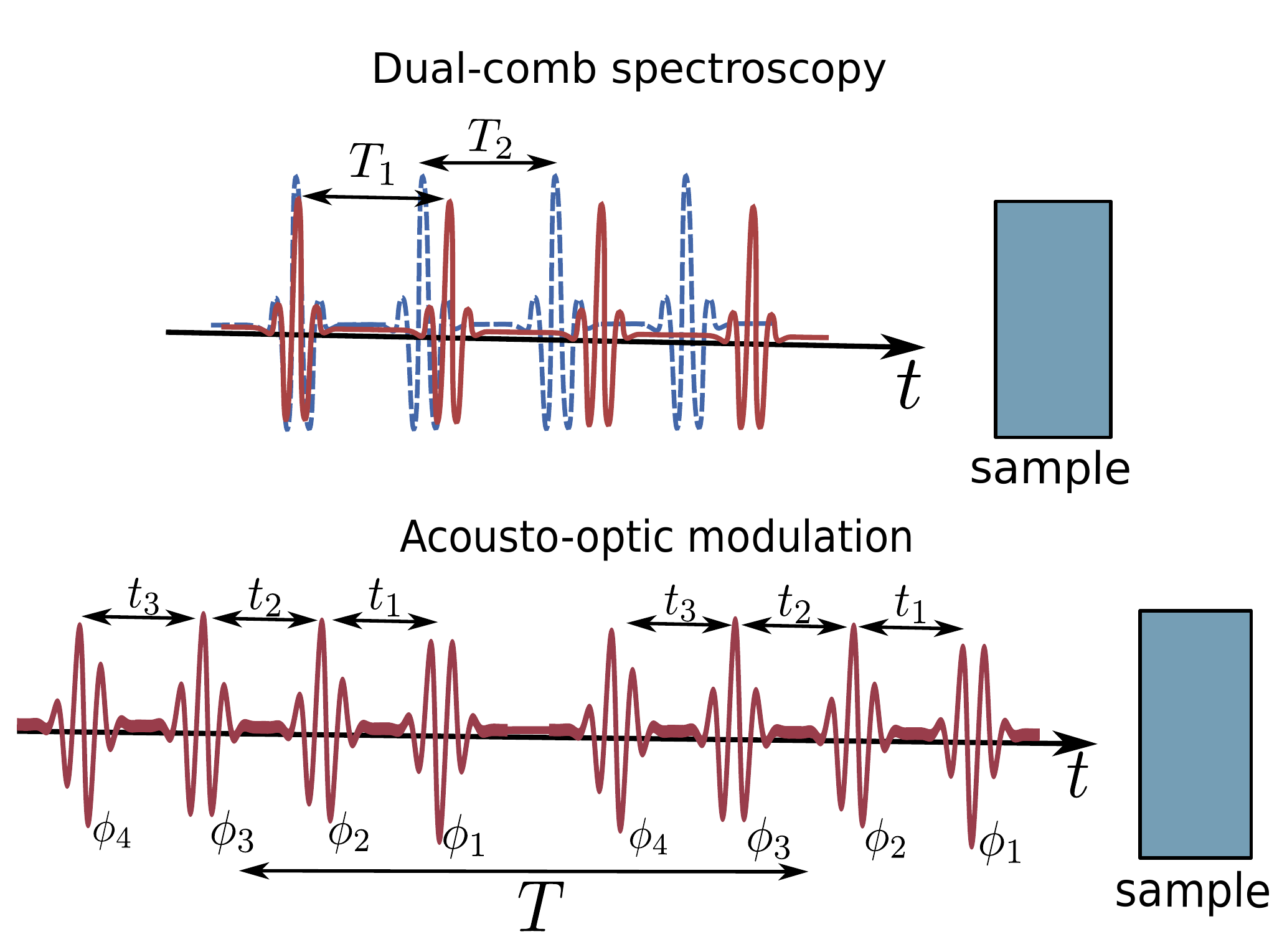}
\caption{Top : Two combs with different repetition frequency (Eq.\ (\ref{eq:Eomega})) are overlaid and interact with the sample. Different interactions pathways are modulated with different slow frequencies. Bottom : multiple pulses with an AOM frequency $\phi_i$ are combined with a varying pulse delay. By accumulating over multiple sequences, interactions pathways are also modulated at slow frequencies.}
 \label{sketch}
\end{figure}

\section{Coherent vs. incoherent signals}\label{sec:signals}

First, we demonstrate that coherent detection can be described as the rate of material energy change due to the laser combs.  
We then discuss incoherent fluorescence detection, which has recently been used to measure two-photon absorption of Rubidium vapor \cite{hipke2014broadband}. We derive expressions for the fluorescence signal, using the excited state population as a proxy quantity for this signal.  In both cases, we derive the signal expressions in terms of the material quantities without specifying how they are generated, thus subsuming arbitrary-order interaction and allowing specialization to linear or nonlinear signals in sections III and IV.  

\subsection{Coherent Signal; Broadband Nonlinear Transmission}\label{sec:transderiv}

The time-dependent dissipation of field energy by the interaction with matter can be measured as the nonlinear transmission.  We start with a model Hamiltonian given by the sum of a material-only term $\hat{H}_0$ and a dipolar term coupling the electric field $E(t)$ to the matter
\begin{align}
\hat{H}=\hat{H}_0-E(t)\hat{V}
\end{align}
where $\hat{V}$ is the dipole operator.  We define the signal as the rate of change of material energy   
\begin{align}
S(t)\equiv\frac{d}{dt}\langle\hat{H}\rangle(t)=\frac{d}{dt}\text{Tr}\left[\hat{H}(t)\rho(t)\right]=\text{Tr}\left[\dot{\hat{H}}(t)\rho(t)+\hat{H}(t)\dot{\rho}(t))\right]
\end{align}
Working in the Schr\"odinger picture and using the Liouville equation $\dot{\rho}=-i[\hat{H},\rho]$ then gives 
\begin{align}
S(t)=\text{Tr}\left[\dot{\hat{H}}(t)\rho(t)\right]-i\text{Tr}\left[\hat{H}^2(t)\rho(t)-\hat{H}(t)\rho(t))\hat{H}(t)\right]=\langle\frac{d}{dt}\hat{H}\rangle=-\dot{E}(t)\langle\hat{V}\rangle(t)
\end{align}
 where we have used the invariance of the trace operation to cyclic permutations and the fact that only the electric field depends explicitly on time. The total energy change due to the field, given by the time integration
\begin{align}
\Delta H(t)\equiv\int_{-\infty}^td\tau\frac{d}{d\tau}\langle\hat{H}\rangle,
\end{align}
can thus be found by substituting the Fourier transforms of the electric field and dipole expectation value
\begin{align}
&\dot{E}(t)=\frac{d}{dt}\int d\omega E(\omega)e^{-i\omega t}=-i\int d\omega E(\omega)\omega e^{-i\omega t}\\ \notag
&\langle\hat{V}\rangle(t)=\int d\omega \langle\hat{V}\rangle(\omega)e^{-i\omega t}
\end{align}
to obtain
\begin{align}\label{eq:DeltaH5}
\Delta H(t)=\frac{i}{(2\pi)^2}\int_{-\infty}^td\tau\int d\omega_1\omega_1\int d\omega_2E(\omega_1)\langle \hat{V}\rangle(\omega_2)e^{-i(\omega_1+\omega_2)\tau}
\end{align}
In the long-time limit ($t\to\infty$), the time-integration gives a $\delta$-function in frequencies and the total dissipated energy is
\begin{align}\label{eq:DeltaH}
\Delta H=\frac{-i}{(2\pi)^2}\int d\omega E(-\omega)\langle\hat{V}\rangle(\omega)\omega=\frac{1}{(2\pi)^2}\Im\int d\omega E^*(\omega)\langle\hat{V}\rangle(\omega)\omega,
\end{align}
where the last equality follows from $E^*(\omega)=E(-\omega)$ and the fact that $\Delta H$ is real. 
This reproduces the standard starting point for arbitrary-order nonlinear spectroscopic signals.  
To proceed further we must specify the electric field and some care must be taken when considering one which is temporally unlimited.  
The simplest example of such a case is a continuous wave $E$-field but a continuously applied pulsed laser is also temporally unlimited.  In such cases, Eq.\ (\ref{eq:DeltaH}) diverges and the material dissipates an infinite total amount of energy.  The meaningful observable should then be the rate of this dissipation
\begin{align}\label{eq:dDeltaHT}
S(t)=-\dot{E}(t)\langle\hat{V}\rangle(t)=\frac{-1}{(2\pi)^2}\Im\int d\omega_1\omega_1\int d\omega_2E(\omega_1)\langle \hat{V}\rangle(\omega_2)e^{-i(\omega_1+\omega_2)t}
\end{align}
or its Fourier transform
\begin{align}\label{eq:dDeltaHomega}
S(\omega)=\frac{-1}{2\pi}\Im\int d\omega_1\omega_1d\omega_2E(\omega_1)\langle\hat{V}\rangle(\omega_2)\delta(\omega-\omega_1-\omega_2)= \frac{-1}{2\pi}\Im\int d\omega'(\omega-\omega')E(\omega-\omega')\langle\hat{V}\rangle(\omega')
\end{align}
which is the rate of energy dissipation as a function of frequency.
In the applications given hereafter we consider infinite combs. We thus use Eqs.\  (\ref{eq:dDeltaHT})  and (\ref{eq:dDeltaHomega})  rather than (\ref{eq:DeltaH5}) and (\ref{eq:DeltaH}). We then analyze the temporal modulation
of the signals for various comb configurations.

\subsection{Incoherent, Fluorescence-Detected, Signals}\label{sec:fluorderiv}
So far, we have discussed a coherent detection of the gain (or loss) of material energy due to the field. Alternatively, one can detect the time-dependent population $\hat{P}_e\equiv\vert e\rangle \langle e\vert$ of some selected excited state $e$, where $\hat{P}$ is a projection operator.  
While multi-frequency comb spectroscopy has traditionally been done in a heterodyne fashion, we demonstrate below that the same ideas can be imported into fluorescence detection schemes.  By exciting the system with a field composed of multiple combs, the same phase-modulation of the signal occurs in fluorescence as in the heterodyne case.  As before, when the field is not temporally confined, the physically relevant signal is the rate-of-change of this population.  Fluorescence is a readily identifiable proxy-signal for this quantity.  The signal is then proportional to the time-dependent population flux of an emitting excited state
\begin{align}
\frac{d}{dt}\langle\hat{P}_e\rangle(t)\equiv S_{e}(t)=\text{Tr}\left[\dot{\hat{P}}_e\rho+\hat{P}_e\dot{\rho}\right]=i\text{Tr} \left[\hat{P}_e\hat{H}\rho-\hat{P}_e\rho\hat{H}\right],
\label{sigdef}
\end{align}
where we have used the Liouville equation and $\dot{\hat{P}}_e=0$ when $\vert e\rangle$ is an eigenstate of $\hat{H}_0$.  This can be further simplified by making use of the cyclic invariance of the trace and evaluating the resulting commutator
\begin{align}
S_e(t)=i\text{Tr}\left[[\hat{P}_e,H]\rho\right]=\Im\left[E(t)\langle\hat{P}_e\hat{V}\rangle(t)\right]=\Im\left[\int d\omega_1d\omega_2 E(\omega_1)\langle\hat{P}_e\hat{V}\rangle(\omega_2)e^{i(\omega_1+\omega_2)t}\right]
\end{align}
or, in the frequency domain,
\begin{align}
S_e(\omega)=\Im\left[\int d\omega_1d\omega_2E(\omega_1)\langle\hat{P}_e\hat{V}\rangle(\omega_2)\delta(\omega-\omega_1-\omega_2)\right]= \Im\left[\int d\omega' E(\omega-\omega')\langle\hat{P}_e\hat{V}\rangle(\omega')\right].
\end{align}
These expressions for the incoherent population signal are analogous to Eqs.\ (\ref{eq:dDeltaHT}) and (\ref{eq:dDeltaHomega}) for the rate of energy gain/loss and may be similarly expanded order-by-order. The only difference between the two types of detection is that the material operator of relevance is now the projected dipole product $\hat{P}_e\hat{V}$ rather than the entire dipole $\hat{V}$.  If we define the projected susceptibilities 
\begin{align}
\langle\hat{P}_e\hat{V}\rangle^{(1)}(\omega)&\equiv\int d\omega' E(\omega')\chi^{(1)}_e(-\omega,\omega')\delta(-\omega+\omega')\\ \notag
\langle\hat{P}_e\hat{V}\rangle^{(3)}(\omega)&\equiv\int d\omega_1d\omega_2d\omega_3E(\omega_3)E(\omega_2)E(\omega_1) \chi^{(3)}_e(-\omega,\omega_3,\omega_2,\omega_1)\delta(-\omega+\omega_3+\omega_2+\omega_1)
\end{align}
then all results derived formally for the coherent transmission signal will apply immediately to these incoherent signals by the simple substitution $\chi\to\chi_e$.  
We note that, since the projection operators satisfy $\sum\hat{P}_e+\hat{P}_g=1$, we have $\sum\chi_e+\chi_g=\chi$.  However, the $\chi_g$ signal obviously does not contribute to a fluorescence signal since it is the probability of returning to the ground state after the last dipole interaction.  
\par
The projected susceptibility can also be written as a Liouville-space superoperator expression
\begin{align}
\chi^{(n)}_e(-\omega,\omega_n,\dots,\omega_1)\equiv\langle\langle\mathbbm{1}\vert\hat{P}_{e,L}\hat{V}_L(\omega)\hat{V}_-(\omega_n)\dots\hat{V}_-(\omega_1)\vert\rho(-\infty)\rangle\rangle,
\end{align}
where $\langle\langle\mathbbm{1}\vert$ is the trace operator.  The subscript ``$L$" indicates  action on the left in Hilbert space $\hat{O}_L\vert\rho\rangle\rangle\leftrightarrow\hat{O}\rho$ while the ``$-$" subscript stands for the commutator $\hat{O}_-\vert\rho\rangle\rangle\leftrightarrow\hat{O}\rho-\rho\hat{O}$.  We could also simplify this by acting with the projection operator to the left on the trace operator $\langle\langle\mathbbm{1}\vert\hat{P}_{e,L}=\langle\langle ee\vert$
\begin{align}
\chi^{(n)}_e(-\omega,\omega_n,\dots,\omega_1)\equiv\langle\langle ee\vert\hat{V}_L(\omega)\hat{V}_-(\omega_n)\dots\hat{V}_-(\omega_1)\vert gg\rangle\rangle,
\label{chien}
\end{align}
where we have assumed the equilibrium density matrix to be the ground state.  For comparison, the superoperator expression for the full $\chi^{(n)}$ is 
\begin{align}
\chi^{(n)}(-\omega,\omega_n,\dots,\omega_1)\equiv\langle\langle\mathbbm{1}\vert\hat{V}_L(\omega)\hat{V}_-(\omega_n)\dots\hat{V}_-(\omega_1)\vert gg\rangle\rangle.
\label{chien2}
\end{align}

\section{Dual Frequency Comb Spectroscopy}
\label{sec:expand}
\subsection{Time domain derivation}
We now apply the results of section \ref{sec:signals} to frequency combs as shown at the top of Fig.\ \ref{sketch}. 
We first consider the linear response of a system to a pair of frequency combs.  
The time-domain electric field of a pulse train corresponds to a comb in frequency space so that, for the $j$-th pulse train with $T_j$ the repetition period of the mode-locked pulses, we have \cite{cundiff2002phase}
\begin{align}\label{pulsedef}
E_j(t)=\sum_n \tilde{E}(t-nT_j)e^{i(\omega_c t-n\omega_cT_j+n\phi_{\text{ce},j}+\phi_j)}\to E_j(\omega)&=e^{i\phi_j}\tilde{E}(\omega-\omega_c)\sum_ne^{in(\phi_{\text{ce},j}-\omega T_j)}\\ \notag
&=e^{i\phi_j}\tilde{E}(\omega-\omega_c)\sum_n\delta(\omega-n\Delta\omega_j-\omega_{\text{ce},j}),
\end{align}
where we have separated the electric field into an envelope function $\tilde{E}$ and a carrier wave with frequency $\omega_c$,  $\Delta\omega_j\equiv2\pi/T_j$, $j=1,2$ are the laser repetition frequencies, $\phi_{\text{ce},j}$ is the pulse-to-pulse carrier envelope phase shift leading to the uniform frequency shift $\omega_{\text{ce},j}=\phi_{\text{ce},j}\Delta\omega_j$ and $\phi_j$ is an overall phase. This definition can also be used for a pulse series of ultrafast lasers. When an AOM with frequency $\phi_{jn}$ is used, an extra $\phi_{jn} T_j$ is added to the phase. The ability to measure $\phi_\text{ce}$ for optical frequency combs, by self referencing for example, was instrumental in allowing their use for high-precision metrology since it determines the mapping betweek optical frequencies $\omega$ and RF frequencies $\Delta \omega$ and $\omega_\text{ce}$ \cite{cundiff2003colloquium}.  In dual- and multi-comb spectroscopies, active control of $\phi_\text{ce}$ allows a much higher degree of accuracy \cite{coddington2016dual} and phase-sensitive schemes can be used to separate linear from nonlinear signals \cite{lomsadze2017four}. While knowledge of $\omega_{\text{ce}}$ is thus necessary to precisely determine the frequency of the comb teeth, it is not essential for understanding the key property of downshifting high frequency material response, where the downshift factor depends on the laser repetition rate difference. As the focus of this paper is the derivation of general expressions for AOM and multi-comb nonlinear spectroscopy, we do not further address these important details in this manuscript and, for simplicity, take $\omega_{\text{ce},j}$ and $\phi_j$ to be zero with the understanding that they can be easily resurrected at any point as needed.

We consider a DFC composed of two frequency combs $E=\sum_jE_j$. We assume that the two combs have the same envelope function and are in phase so that the DFC is given by
\begin{align}\label{eq:Eomega}
E(\omega)=\tilde{E}(\omega-\omega_c)\sum_n\left(\delta(\omega-n\Delta\omega_1)+\delta(\omega-n\Delta\omega_2)\right).
\end{align}
In practical experimental implementations\cite{hipke2014broadband}, the repetition rates are chosen such that 
\begin{align}\label{eq:DeltaCond}
\frac{\Delta\omega_2-\Delta\omega_1}{\Delta\omega_{1(2)}}\equiv\frac{\delta\omega_2}{\Delta\omega_{1(2)}}\ll1
\end{align}
For this reason, we will use the notation $\Delta\omega_1\equiv\Delta\omega$, $\Delta\omega_2=\Delta\omega+\delta\omega_2$ which is readily generalizable to $\Delta\omega_j=\Delta\omega+\delta\omega_j$ where we set $\delta\omega_1=0$ without loss of generality. 
\par
We first consider the linear response in which the dipole expectation value is given by
\begin{align}\label{eq:Vomega}
\langle \hat{V}\rangle(\omega)=\int d\omega' \chi^{(1)}(-\omega,\omega')E(\omega')\delta(-\omega+\omega')= \chi^{(1)}(-\omega,\omega)E(\omega).
\end{align}
where $\chi^{(1)}$ is the linear susceptibility.  Upon substituting Eqs.\ (\ref{eq:Eomega}) and (\ref{eq:Vomega}) into Eq.\ (\ref{eq:dDeltaHT}), we find that the time-dependent rate of energy change has four contributions, which we denote by $S_{jk}(t)$, $j,k=1,2$, depending on whether the material interacts with comb 1 or 2 in each $E$ field.  The ``diagonal" terms come as  
\begin{align}
S^{(1)}_{jj}(t)=&-\Im\sum_{nm}\frac{n\Delta\omega_j}{(2\pi)^2}\tilde{E}(n\Delta\omega_j-\omega_c)\tilde{E}(m\Delta\omega_j-\omega_c)\chi^{(1)}(-m\Delta\omega_j,m\Delta\omega_j)e^{-i(n-m)\Delta\omega_jt}
\end{align}
while the cross terms are
\begin{align}\label{eq:DeltaH12}
S^{(1)}_{12}(t)=&-\Im\sum_{nm}\frac{n\Delta\omega_1}{(2\pi)^2}\tilde{E}(n\Delta\omega_1-\omega_c)\tilde{E}(m\Delta\omega_2-\omega_c)\chi^{(1)}(-m\Delta\omega_2,m\Delta\omega_2)e^{-i(n-m)\Delta\omega t}e^{-im(\delta\omega_2-\delta\omega_1) t}
\end{align}
with $\Delta H_{21}(t)$ following similarly by $1\leftrightarrow 2$.  The time modulation represented by the exponential factors is critically important.  
Note that the diagonal terms can come modulated with any multiple of the repetition frequencies while the cross terms come modulated with a multiple of one of the repetition frequencies and a $\delta\omega$ shift.  
The double summation can be simplified by considering the time-dependent signal that would result from a low-pass filtering where the cut-off frequency is less than either $\Delta\omega_{j}$.  This would eliminate all terms $n\ne m$, rendering $\Delta H_{jj}$ a static, DC contribution.  The cross terms however, would then come as
\begin{align}
S^{(1)}_{12}(t)=-\Im\sum_{n}\frac{n\Delta\omega}{(2\pi)^2}\tilde{E}(n\Delta\omega-\omega_c)\tilde{E}(n\Delta\omega_2-\omega_c)\chi^{(1)}(-n\Delta\omega_2,n\Delta\omega_2)e^{-in\delta\omega t}.
\end{align}
where we have set $\delta\omega_2=\delta\omega$ as it is the only such term at linear order.  The $n$th term in this summation carries the information of the material response at $n\Delta\omega_{2}$ but is modulated in time by the factor $e^{in\delta\omega t}$, i.e., at a  down-shifted frequency of $\delta\omega$.  Thus, in a temporal detection of the rate of energy dissipated (absorbed) by the matter, the response at an optical frequency $n\Delta\omega_2$ will be modulated at a frequency $n\delta\omega$.  Since the externally-controlled ratio between these two factors can reach $\frac{\delta\omega}{\Delta\omega}\approx 10^{-6}$, this effectively down-shifts the optical frequency to the microwave or even RF regime, where it is easily detected in the time-domain by standard electronics. The material susceptibility at an optical frequency is accompanied by a microwave oscillation.

\subsection{Frequency Domain Derivation}
Equivalently, we can begin with the general expression (Eq.\ (\ref{eq:dDeltaHomega})) for the frequency-dependent rate of energy dissipation
\begin{align}
S(\omega)= \frac{-1}{2\pi}\Im\int d\omega'(\omega-\omega')E(\omega-\omega')\langle\hat{V}\rangle(\omega')
\end{align}
Inserting Eq.\ (\ref{eq:Eomega}) for the electric field gives
\begin{align}
S(\omega)= \frac{-1}{2\pi}\Im\sum_{j=1,2}\sum_nn\Delta\omega_j\tilde{E}(n\Delta\omega_j-\omega_c)\langle\hat{V}\rangle(\omega-n\Delta\omega_j)
\end{align}
The linear response is then obtained by inserting Eq.\ (\ref{eq:Vomega}) which gives
\begin{align}
S^{(1)}(\omega)=& \frac{-1}{2\pi}\Im\sum_{j=1,2}\sum_nn\Delta\omega_j\tilde{E}(n\Delta\omega_j-\omega_c)\chi^{(1)}(n\Delta\omega_j-\omega,\omega-n\Delta\omega_j)E(\omega-n\Delta\omega_j).
\end{align}
Inserting the form of the $E$ field then results in  
\begin{align}
S^{(1)}(\omega)&=\frac{-1}{2\pi}\Im\sum_{j,k=1,2}\sum_{nm}n\Delta\omega_j\tilde{E}(n\Delta\omega_j-\omega_c)\tilde{E}(m\Delta\omega_k-\omega_c)
\chi^{(1)}(-m\Delta\omega_k,m\Delta\omega_k)\delta(n\Delta\omega_{j}+m\Delta\omega_k-\omega)
\end{align}
where we have used the $\delta$-function to simplify the arguments of $\chi^{(1)}$. Upon setting $j=1$, $k=2$, we recover the Fourier transform of Eq.\ (\ref{eq:DeltaH12}).  We can also recast this in an alternative form by relabelling $n+m=n'$
\begin{align}
S^{(1)}(\omega)&=\frac{-1}{2\pi}\Im\sum_{j,k=1,2}\sum_{m}(\omega-m\Delta\omega_k)\tilde{E}(\omega-m\Delta\omega_k-\omega_c)\tilde{E}(m\Delta\omega_k-\omega_c)\\ \notag
&\times\chi^{(1)}(-m\Delta\omega_k,m\Delta\omega_k)\sum_{n'}\delta(n'\Delta\omega+m(\delta\omega_k-\delta\omega_j) -\omega)
\end{align}


Here the relabelling (and the use of the $\delta$-function) has allowed us to put one of the summations purely on the $\delta$-function term, into which we have substituted the definition of the $\Delta\omega_{j,k}$.  This makes clear that, for each choice of $n'$, there will be a comb with spikes separated by $\delta\omega_k-\delta\omega_j$ corresponding to different values of $m$ and each of these separate combs will be shifted by $n'\Delta\omega$.  Finally, we may consider a low-pass filtered signal as before to select the $n'=0$ term and obtain the simple formula
\begin{align}
S^{(1)}(\omega)&=\frac{1}{2\pi}\Im\sum_{j,k=1,2}\sum_{m}m\Delta\omega_j\tilde{E}(-m\Delta\omega_j-\omega_c)\tilde{E}(m\Delta\omega_k-\omega_c)\chi^{(1)}(-m\Delta\omega_k,m\Delta\omega_k)\delta(m(\delta\omega_k-\delta\omega_j) -\omega)
\end{align}
It is then immediately clear that the $j=k$ terms will all contribute a DC component at $\omega=0$ while the $j=1$, $k=2$ term will generate spikes at $\omega=m\delta\omega_2$, the $j=2$, $k=1$ term generates spikes at $\omega=-m\delta\omega_2$ which overlap since $m$ takes positive and negative integer values.
The above frequency-domain approach is clearly equivalent to the time-domain approach of the previous section but is easier to visualize because of the explicit appearance of the $\delta$-function in the frequency domain.  
 This expression can be simplified further by assuming that only one comb passes through the material while the other is used for the heterodyne detection so that, setting $j=2,k=1$, we obtain
\begin{align}
S^{(1)}_{21}(\omega)&=\frac{1}{2\pi}\Im\sum_{m}m\Delta\omega_2\tilde{E}(-m\Delta\omega_2-\omega_c)\tilde{E}(m\Delta\omega-\omega_c)\chi^{(1)}(-m\Delta\omega,m\Delta\omega)\delta(-m\delta\omega -\omega)
\end{align}
where, for simplicity, we have relabelled $\delta\omega_2=\Delta\omega_2-\Delta\omega=\delta\omega$ since this is the only such quantity in the linear case.  Similarly, detecting the reference beam corresponds to setting $j=1,k=2$ to obtain
\begin{align}
S^{(1)}_{12}(\omega)&=\frac{1}{2\pi}\Im\sum_{m}m\Delta\omega\tilde{E}(-m\Delta\omega-\omega_c)\tilde{E}(m\Delta\omega_2-\omega_c)\chi^{(1)}(-m\Delta\omega_2,m\Delta\omega_2)\delta(m\delta\omega -\omega)
\end{align}

\section{Quad Comb Spectroscopy; measuring The Third-Order Response}\label{sec:4comb}

Extending the previous formalism to any nonlinear response is straightforward.  Here, we calculate the third-order signals obtained by the application of 4 combs.  Third-order signals measured using 2 or 3 combs are then special cases while the use of higher numbers of combs would simply be a sum over possible 4-sets of the applied combs.  A recent implementation \cite{lomsadze2017four} of a four-wave mixing measurement effectively utilizes 3 combs by starting with 2, splitting one of these, and using an AOM in one path of the apparatus.  In that work, a phase cancellation scheme is utilized to separate out the desired signal.  While the details of experimental implementation of such techniques are not completely straightforward, the advantages discussed above (short acquisition times at high resolution) justify their use.  

Generalizing Eq.\ (\ref{eq:Eomega}), we write the electric field as
\begin{align}\label{eq:4combfield}
E(\omega)=\tilde{E}(\omega-\omega_c)\sum_{j=\lbrace 1,2,3,4\rbrace}\sum_n \delta(\omega-n\omega_j)
\end{align}
and the dipole expectation value at this order is
\begin{align}
\langle\hat{V}\rangle(\omega)=\int d\omega_1d\omega_2d\omega_3E(\omega_1)E(\omega_2)E(\omega_3) \chi^{(3)}(-\omega,\omega_3,\omega_2,\omega_1)\delta(-\omega+\omega_3+\omega_2+\omega_1)
\end{align}
which, upon substitution of Eq.\ (\ref{eq:4combfield}), becomes
\begin{align}
S^{(3)}(\omega)=&\frac{-1}{2\pi}\Im\sum_{j_1,j_2,j_3, j_4}\sum_{nmpq} n\Delta\omega_{j_1}\tilde{E}(n\Delta\omega_{j_1}-\omega_c)\tilde{E}(m\Delta\omega_{j_2}-\omega_c)\tilde{E}(p\Delta\omega_{3}-\omega_c)\tilde{E}(q\Delta\omega_{j_4}-\omega_c)\\ \notag
&\times\chi^{(3)}(n\Delta\omega_{j_1}-\omega,m\Delta\omega_{j_2},p\Delta\omega_{j_3},q\Delta\omega_{j_4}) \delta(n\Delta\omega_{j_1}-\omega+m\Delta\omega_{j_2}+ p\Delta\omega_{j_3}+q\Delta\omega_{j_4}).
\end{align}
where each of the $j_1,\dots j_4$ are summed over $\lbrace 1,\dots 4\rbrace$ signifying the 4 applied combs.
To simplify this expression, we first define $n+m+p+q\equiv n'$ and $\omega_{mpq}\equiv m\Delta\omega_{j_2}+ p\Delta\omega_{j_3}+q\Delta\omega_{j_4} $ which gives
\begin{align}
S^{(3)}(\omega)=&\frac{-1}{2\pi}\Im\sum_{j_1,j_2,j_3, j_4}\sum_{mpq}(\omega-\omega_{mpq})\tilde{E}(\omega-\omega_{mpq}-\omega_c)\tilde{E}(m\Delta\omega_{j_2}-\omega_c)\tilde{E}(p\Delta\omega_{3}-\omega_c)\tilde{E}(q\Delta\omega_{j_4}-\omega_c)\\ \notag
&\times\chi^{(3)}(-\omega_{mpq},m\Delta\omega_{j_2},p\Delta\omega_{j_3},q\Delta\omega_{j_4}) \sum_{n'}\delta(-\omega+n'\Delta\omega_{j_1}+m(\delta\omega_{j_2}-\delta\omega_{j_1}) +p(\delta\omega_{j_3}-\delta\omega_{j_1})+q(\delta\omega_{j_4}-\delta\omega_{j_1}))
\end{align}
where all high-frequency components are at multiples of $n'$.  Applying a low-pass filter, as in the linear case, thus finally selects the $n'=0$ term
\begin{align}
S^{(3)}(\omega)=&\frac{1}{2\pi}\Im\sum_{j_1,j_2,j_3, j_4}\sum_{mpq}(m+p+q)\Delta\omega_{j_1}\tilde{E}(-(m+p+q)\Delta\omega_{j_1}-\omega_c)\tilde{E}(m\Delta\omega_{j_2}-\omega_c)\tilde{E}(p\Delta\omega_{3}-\omega_c)\tilde{E}(q\Delta\omega_{j_4}-\omega_c)\\ \notag
&\times\chi^{(3)}(-\omega_{mpq},m\Delta\omega_{j_2},p\Delta\omega_{j_3},q\Delta\omega_{j_4}) \delta(-\omega+m(\delta\omega_{j_2}-\delta\omega_{j_1}) +p(\delta\omega_{j_3}-\delta\omega_{j_1})+q(\delta\omega_{j_4}-\delta\omega_{j_1})).
\end{align}
where the first argument of $\chi^{(3)}$ can also be written as $-\omega_{mpq}=-(m+p+q)\Delta\omega_{j_1}-\omega$ via the $\delta$-function. Setting the detected comb to be the reference comb then gives $j_1=1$ and the signal corresponds to
\begin{align}
S^{(3)}_1(\omega)=&\frac{1}{2\pi}\Im\sum_{mpq}(m+p+q)\Delta\omega\tilde{E}(-(m+p+q)\Delta\omega-\omega_c)\tilde{E}(m\Delta\omega_{j}-\omega_c)\tilde{E}(p\Delta\omega_{3}-\omega_c)\tilde{E}(q\Delta\omega_{4}-\omega_c)\\ \notag
&\times\chi^{(3)}(-(m+p+q)\Delta\omega-\omega,m\Delta\omega_{2},p\Delta\omega_{3},q\Delta\omega_{4}) \delta(-\omega+m\delta\omega_{2} +p\delta\omega_{3}+q\delta\omega_{4}).
\end{align}
where we have used the symmetry of $\chi$ to eliminate the summations over the $j_i$ (e.g., the transformation $(j_2,m)\leftrightarrow (j_3,p)$ yields the same expression).  
The presence of the integer summations means that the various frequency dimensions of the nonlinear response are essentially ``folded" into the single observation axis $\omega$, preventing an immediate recovery of the full $\chi$ at precise frequencies. By varying $\delta\omega_j$, we can then produce a 4-dimensional signal $S(\omega,\delta\omega_2,\delta\omega_3,\delta\omega_4)$.  The summations over $mpq$ can be thought of as tensor contractions (of the tensors represented by the $\delta$-function and the product of field envelopes with $\chi^{(3)}$) and a generalized tensor equation inversion procedure can then be used to determine $\chi^{(3)}$.  The essential point remains the same as in the case of linear spectroscopy, i.e., all measurements are conducted at the frequency scale determined by the $\delta\omega$ but they are sampling the material response $\chi$ at the $\Delta\omega$-scale, which may be 5 or 6 orders of magnitude higher in frequency.

\subsection{Dual-Comb Third-Order Techniques}
The special case of dual-comb, third-order spectroscopy in which all perturbative interactions are with comb 2 and the signal is heterodyne detected with respect to comb 1 gives
\begin{align}
S^{(3)}_1(\omega)=&\frac{1}{2\pi}\Im\sum_{mpq}(m+p+q)\Delta\omega\tilde{E}(-(m+p+q)\Delta\omega-\omega_c)\tilde{E}(m\Delta\omega_{2}-\omega_c)\tilde{E}(p\Delta\omega_{2}-\omega_c)\tilde{E}(q\Delta\omega_{2}-\omega_c)\\ \notag
&\times\chi^{(3)}(-(m+p+q)\Delta\omega_2,m\Delta\omega_{2},p\Delta\omega_{2},q\Delta\omega_{2}) \delta(-\omega+(m+p+q)\delta\omega).
\end{align}
It is immediately clear that the third-order DFC technique does not provide enough tunable parameters to allow for a full inversion to obtain $\chi^{(3)}$ due to the dimensionality.
Another special case is two interactions with each of two combs which becomes
\begin{align}
S^{(3)}_1(\omega)=&\frac{1}{2\pi}\Im\sum_{j_2,j_3, j_4}\sum_{mpq}(m+p+q)\Delta\omega\tilde{E}(-(m+p+q)\Delta\omega-\omega_c)\bigg[\tilde{E}(m\Delta\omega-\omega_c)\tilde{E}(p\Delta\omega_{2}-\omega_c)\tilde{E}(q\Delta\omega_2-\omega_c)\\ \notag
&\times\chi^{(3)}(-(m+p+q)\Delta\omega-\omega,m\Delta\omega,p\Delta\omega_{2},q\Delta\omega_{2}) \delta(-\omega+(p+q)\delta\omega)+\tilde{E}(m\Delta\omega_2-\omega_c)\tilde{E}(p\Delta\omega-\omega_c)\tilde{E}(q\Delta\omega_{2}-\omega_c)\\ \notag
&\times\chi^{(3)}(-(m+p+q)\Delta\omega-\omega,m\Delta\omega_2,p\Delta\omega,q\Delta\omega_{2}) \delta(-\omega+(m+q)\delta\omega)+\tilde{E}(m\Delta\omega_{2}-\omega_c)\tilde{E}(p\Delta\omega_{2}-\omega_c)\tilde{E}(q\Delta\omega-\omega_c)\\ \notag
&\times\chi^{(3)}(-(m+p+q)\Delta\omega-\omega,m\Delta\omega_{2},p\Delta\omega_{2},q\Delta\omega) \delta(-\omega+(m+p)\delta\omega)\bigg].
\end{align}
This can be simplified by changing $m\leftrightarrow p$ in the second term and $m\leftrightarrow q$ in the third
\begin{align}
S^{(3)}_1(\omega)=&\frac{1}{2\pi}\Im\sum_{j_2,j_3, j_4}\sum_{mpq}(m+p+q)\Delta\omega\tilde{E}(-(m+p+q)\Delta\omega-\omega_c)\tilde{E}(m\Delta\omega-\omega_c)\tilde{E}(p\Delta\omega_{2}-\omega_c)\\ \notag
&\times \tilde{E}(q\Delta\omega_2-\omega_c)\delta(-\omega+(p+q)\delta\omega)\bigg[\chi^{(3)}(-(m+p+q)\Delta\omega-\omega,m\Delta\omega,p\Delta\omega_{2},q\Delta\omega_{2}) \\ \notag &+\chi^{(3)}(-(m+p+q)\Delta\omega-\omega,p\Delta\omega_2,m\Delta\omega,q\Delta\omega_{2}) \times\chi^{(3)}(-(m+p+q)\Delta\omega-\omega,q\Delta\omega_{2},p\Delta\omega_{2},m\Delta\omega)\bigg].
\end{align}
This expression possesses peaks at every $p'\equiv p+q$ and we thus recast as
\begin{align}
S^{(3)}_1(\omega)=&\frac{1}{2\pi}\Im\sum_{j_2,j_3, j_4}\sum_{mpp'}(m+p')\Delta\omega\tilde{E}(-(m+p')\Delta\omega-\omega_c)\tilde{E}(m\Delta\omega-\omega_c)\tilde{E}(p\Delta\omega_{2}-\omega_c)\\ \notag
&\times \tilde{E}((p'-p)\Delta\omega_2-\omega_c)\delta(-\omega+p'\delta\omega)\bigg[\chi^{(3)}(-(m+p')\Delta\omega-\omega,m\Delta\omega,p\Delta\omega_{2},(p'-p)\Delta\omega_{2}) \\ \notag &+\chi^{(3)}(-(m+p')\Delta\omega-\omega,p\Delta\omega_2,m\Delta\omega,(p'-p)\Delta\omega_{2}) \times\chi^{(3)}(-(m+p')\Delta\omega-\omega,(p'-p)\Delta\omega_{2},p\Delta\omega_{2},m\Delta\omega)\bigg].
\end{align}
We thus see that each choice of $p'$, corresponding to a choice of peak in the detection, fixes the sum of pairs of arguments of $\chi^{(3)}$ as $-p'\Delta\omega_2$ and $p'\Delta\omega_2$.  
The amplitude of this peak is then given by the sum over $m,p$.  
The resulting spectra will reveal resonances whenever $p'\Delta\omega_2$ is a two-photon resonance of the material.  This therefore contains resonances at Raman transitions (frequency differences) as well as two-photon (frequency sum) absorption transitions, depending on the magnitudes of $p'$ and $\Delta\omega$ relative to these material energy scales.  

\section{Time-Domain Comb Spectroscopy; Acousto-Optic Modulation}
\label{section4}
Acousto-optic modulation is an alternative technique that makes use of pulse trains to downshift material responses into detectable frequency regimes (see Fig.\ \ref{sketch}).  
Here, the laser output is separated into 4 paths using multiple beam splitters and delay stages to independently control their arrival time on the sample\cite{doi:10.1063/1.2800560}. Each pulse passes through an AOM that imparts a phase $\phi_j n T$ to the pulse that passes through it ($j = 1,2,3,4$ for each of the 4 pulses).
The $j^\text{th}$ pulse is then given by
\begin{equation}
E_j(t) = \sum_n \tilde{E}_j (t-t_j- n T_j) e^{i\omega(t-t_j - nT) + i\phi_j n T_j}
\end{equation}
where $t_j$ is the time delay of the pulse sequence $j$, $T_j$ is the repetition period of the laser ($\approx$ 1.25$\mu$s), and $n$ is the discrete variable for a pulse within the $j^{\text{th}}$ pulse train.
\par
\begin{figure}[h!]
\centering
\includegraphics[width=0.7\textwidth]{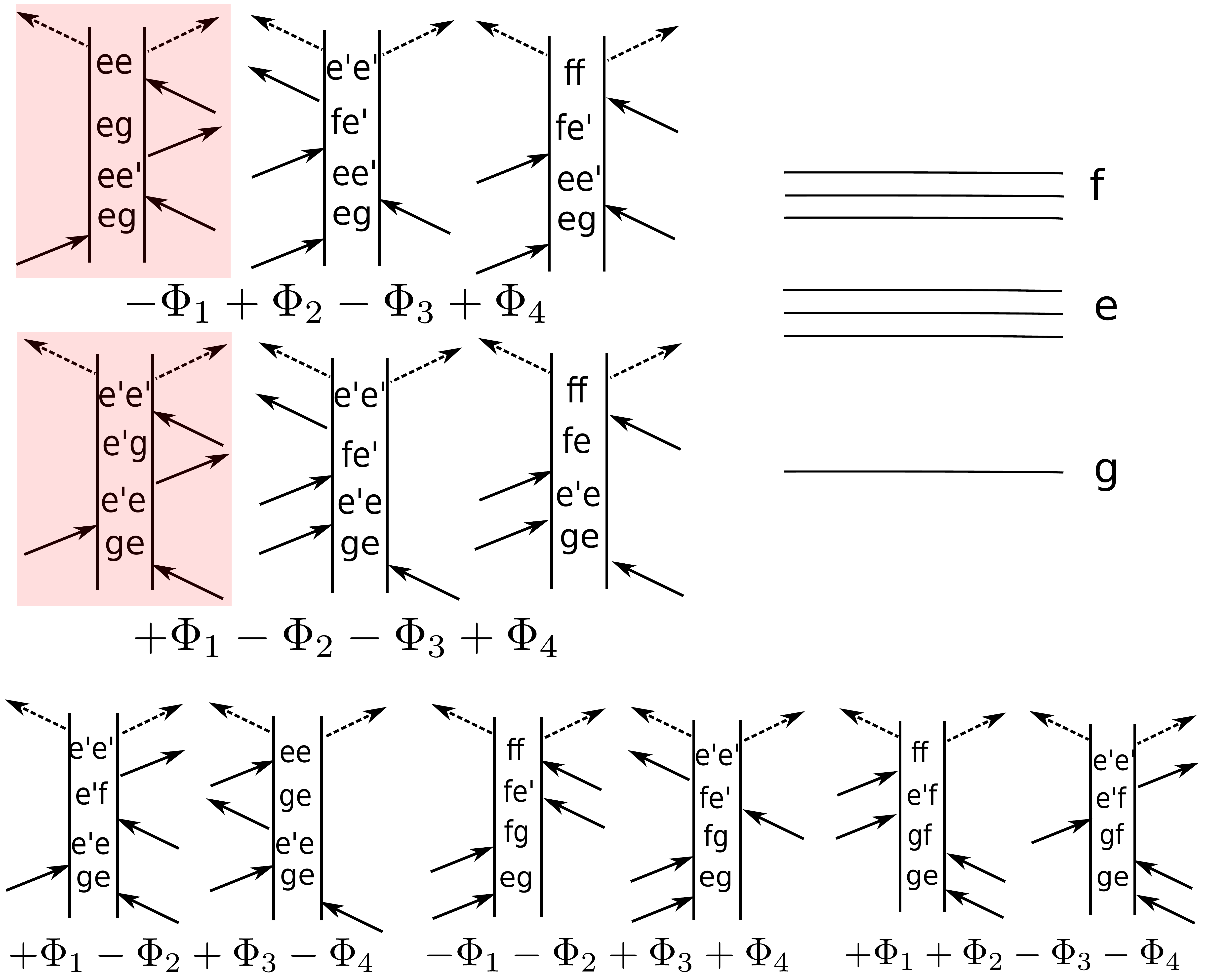}
\caption{Diagrams that contributes to the fluorescence signal are represented with the respective AOM modulation necessary to extract them. Marcus had considered the pathways modulated at $-\Phi_1+\Phi_2-\Phi_3+\Phi_4$ and $+\Phi_1-\Phi_2+\Phi_3-\Phi_4$ and did not consider $f$-manifold excitations. Thus, the two diagrams highlighted in red are the one measured in his experimental scheme.
Other diagrams at the bottom row would require creating different reference waveforms to be measured.}
 \label{diagrams}
\end{figure}

We first assume incoherent detection where the signal is proportional to the time-dependent excited-state population to match existing experiments. Doing so, we do not explicitly consider the emission from the excited state into unpopulated modes of the field. The fluorescence signal is then given, according to Eq.\ (\ref{sigdef}), by
\begin{eqnarray}
S_{e}(t) &=& P_e \ \rho^{(4)}(t)\\
				&=& \int dt_1 dt_2 dt_3 \bold R_e(t_3,t_2,t_1) \cdot \bold E(t)\bold E(t-t_3)\bold E(t-t_3-t_2)\bold E(t-t_3-t_2-t_1)
\end{eqnarray}
where $P_e$ is the projector $|e\rangle\langle e|$ over the detected transition and $\rho^{(4)}(t)$ is the fourth order perturbative expansion on the density matrix. The response function $R(t_3,t_2,t_1)$ is the fourth order correlation function calculated in the usual way and is the Fourier transform of the 4-point matter correlation function defined in Eq.\ (\ref{chien}). It is the sum of various contributions that are represented in Fig.\ \ref{diagrams}.
In the standard convention, each interaction with a leftward-facing arrow brings a $-\phi_j$ frequency shift to the signal while an interaction with a rightward-facing arrow brings a $+\phi_j$ one.\\

Typical values of the frequency differences are $\phi_{43} = \phi_4-\phi_3 = 8$kHz and $\phi_{21} = \phi_2-\phi_1 = 5$kHz \cite{doi:10.1063/1.2800560}.
When the pulses are short enough compared to any material time-scale, the pulses are impulsive and their envelopes $\tilde{E}_j$ can be approximated by Dirac $\delta$-functions and the emitted fluorescence is
\begin{equation}
S(t_{3},t_{2},t_{1},t') = 2 \Re \Big(\sum_i R_i(t_{3},t_{2},t_{1},t') \Big)
\end{equation}
where $t' = mT$ and the sum over $i$ runs over the different interaction pathways\cite{mukbook} presented in the diagrams in Fig.\ \ref{diagrams}. Lock-in amplifiers are used for the detection. The incoming signal is multiplied by a reference waveform and a low pass filter is applied :
\begin{equation}
S_{LI} = \frac{1}{\tau_{LI}} \int_0^{+\infty} dt' S(t_{3},t_{2},t_{1},t') R(t') e^{-t'/\tau_{LI}},
\end{equation}
where $S$ is the detected fluorescence out of the sample, $R$ is a reference waveform created from the incoming pulses and $\tau_{LI}$ is the lock-in low pass time (200 ms). Two different reference waveforms are constructed as follows
\begin{eqnarray}
R_+(t_{3},t_{1}, t') = \cos(\overline{\omega}_{43} t_{3}+\overline{\omega}_{21} t_{1}-(\phi_{43}+\phi_{21})t' -\theta)
\label{Rplus}\\
R_-(t_{3},t_{1}, t') = \cos(\overline{\omega}_{43} t_{3}-\overline{\omega}_{21} t_{1}-(\phi_{43}-\phi_{21})t' -\theta)
\label{Rminus}
\end{eqnarray}

The reference waveforms are created by sending pair of pulses through a monochromator. For pulse 1 and 2 for example, the power density is $|E_1(\omega, t')+E_2(\omega, t')|^2$ and the monochromator is tuned to evaluate the intensity at $\omega=\overline{\omega}_{21}$, leading to Eq.\ (\ref{Rplus}) and Eq.\ (\ref{Rminus}). The extra oscillation in the reference waveform is unnecessary to pick signals modulated at $\Phi_{43}\pm\Phi_{21}$ but it adds an extra oscillation along $t_3$ and $t_1$ that downshifts the frequency of the transition ($e^{-i\omega_{eg}t_3} \longrightarrow e^{-i(\omega_{eg}-\overline{\omega}_{43})t_3}$ and $e^{i\omega_{eg}t_1} \longrightarrow e^{i(\omega_{eg}-\overline{\omega}_{21})t_1}$).
By tuning the monochromator frequency closer to the transition, we can
expect further downconversion. 
In a recent experiment\cite{doi:10.1063/1.2800560}, the monochromator frequency used was $\omega_{\text{m}}$ = 381 THz and the transition frequency observed was $\omega_{\text{eg}}$ = 384 THz to achieve a downshifted frequency of 3 THz. 
This downshifting is used to reduce the number of data points to acquire and to improve the signal to noise ratio.
\\

The recorded signal is split in two and sent to two lock-in amplifiers, using $R_+$ and $R_-$ as a reference wave respectively. $\theta$ is set to 0 or $\pi/2$ to detect the in-phase and in-quadrature component of the signal.
Pathways 1 and 4 are modulated by $\phi_{43}+\phi_{21}$ while pathways 2 and 3 are modulated by  $\phi_{43}+\phi_{21}$. Assuming that the low pass filter removes the oscillatory part of the integrated function, the lock-in $R_+$ extracts the contribution of $R_1+R_4$ to the signal. Similarly, $R_-$ extracts the contribution of $R_2+R_3$.

\begin{figure}[h!]
\centering
\includegraphics[width=0.5\textwidth]{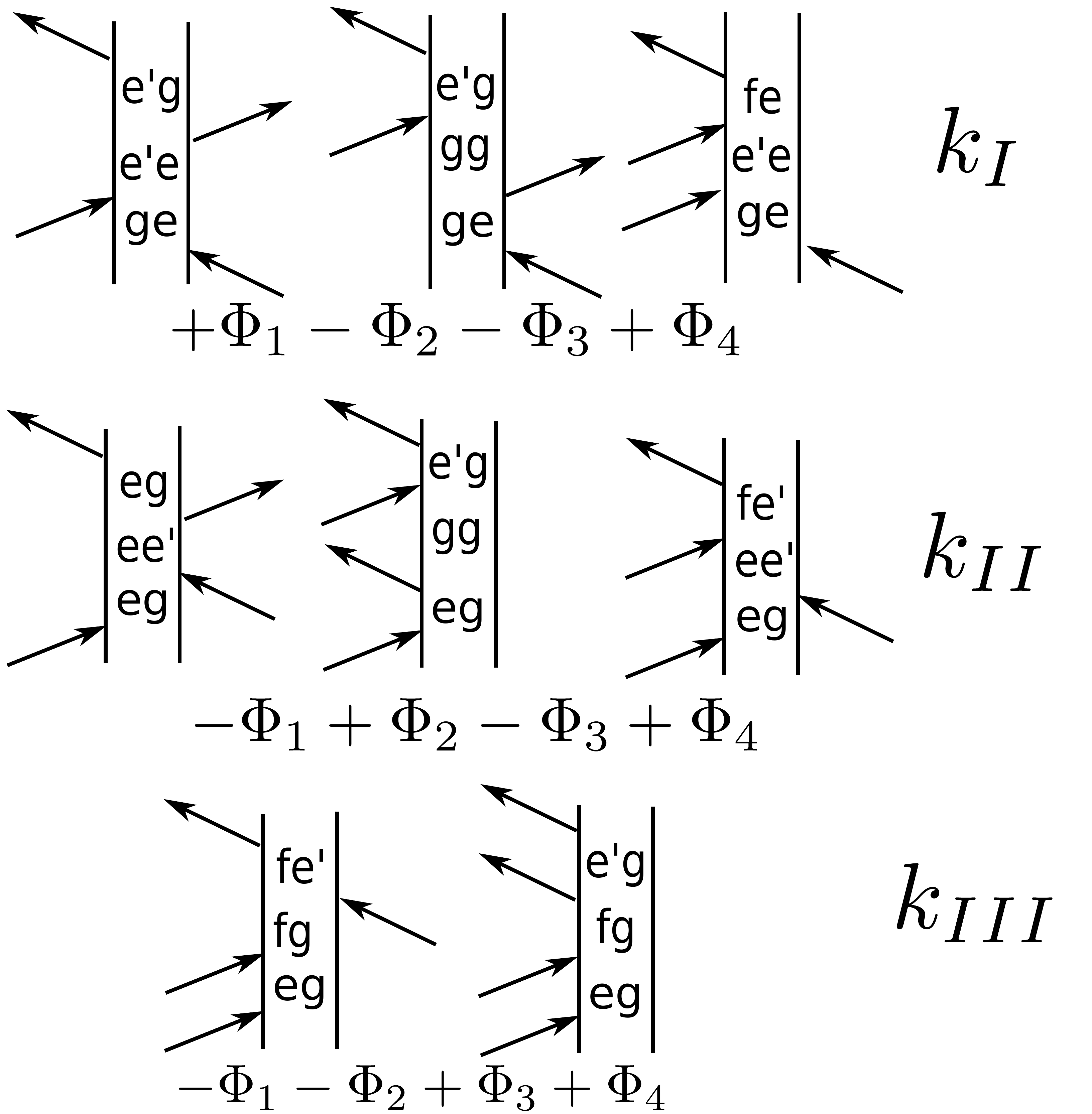}
\caption{Diagrams contributing to heterodyne detected signals for $\bold k_I$, $\bold k_{II}$ and $\bold k_{III}$ techniques.}
 \label{heterodyne}
\end{figure}

We have discussed incoherent fluorescence-detected signals but it is also possible to use heterodyne detection too to measure coherent signals. The heterodyning pulse then carries the frequency $\psi_4$  and a lock-in detection can also be used\cite{borri1999heterodyne}. The corresponding diagrams for the $\bold k_I$, $\bold k_{II}$ and $\bold k_{III}$ techniques are depicted in Fig.\ \ref{heterodyne} and the matter correlation function of Eq.\ (\ref{chien2}) is then measured.

\newpage
\section{Conclusion}

The pulse sequences for the multicomb and the AOM techniques are illustrated in Fig.\ \ref{sketch}.
Dual frequency comb spectroscopy, and its multi-comb generalizations, permit the extraction of a wealth of material information by employing a single pulse sequence.
High-resolution spectral measurements are effectively taken simultaneously at all the comb teeth. 
Thus, data acquisition can be achieved much more rapidly, making the recording less sensitive to fluctuations. 
In dual comb spectroscopy, the signal is modulated by the difference in repetition rates between the two combs, allowing access to high frequency (e.g., optical) material information \textit{via} measurements at much lower (MHz) frequencies with the frequency conversion factor $\delta\omega/\Delta\omega$. 
While phase cancellation schemes can be used to separate the linear from nonlinear signals, the different combinations of comb interactions that can lead to a particular observed frequency become harder to disentangle for multiple frequency combs probing the nonlinear response.  
Tensor equation inversion techniques (see, e.g., \cite{brazell2013solving} and references therein) could be employed to obtain the full multidimensional material susceptibility from varying the comb repetition frequencies relative to each other.  

In multicomb spectroscopies, the frequency down-shifted signal is usually acquired in the time domain and the spectra are recovered by Fourier transforms at the data processing stage. 
In the AOM scheme in contrast, a frequency shift is added in the time domain to each pulse in the sequence that are then accumulated on an intensity detector. This detection scheme carries a Fourier transform experimentally and the frequency to extract different interaction pathways is then selected by a lock-in amplifier.  
Frequency combs and multi-comb techniques have been employed to quickly record a variety of spectroscopic signals at high resolution
 \cite{diddams2000direct,cundiff2003colloquium,schliesser2012mid,diddams2007molecular, hipke2014broadband,ideguchi2012raman,ideguchi2013coherent,lomsadze2017four}. This paper provides a unified analysis for multi-comb spectroscopies that can be used to analyze these experiments and to guide their further development.

We point out that the two techniques have a somewhat blurry separation. Standard ultrafast laser effectively behave like combs when one is able to add an arbitrary phase to each pulse within a sequence. The slow modulation added on interaction pathways with the AOMs also allow to downshift the detected signal.
On the other side, multi-comb spectroscopy can also rely on the use of AOM to add an extra frequency shift to a whole sequence, useful to deconvolute various contributions to the signal (linear from nonlinear or separating different nonlinear interaction pathways) \cite{lomsadze2017four}.
While the two techniques formally overlap, the ability to resolve frequency teeth within the comb allow to do the same acquisition on a much lower timescale, eliminating long time laser fluctuations.

%

\acknowledgements
The support of  the National Science Foundation (grant CHE-1361516) as well as from the Chemical Sciences, Geosciences, and Biosciences division, Office of Basic
Energy Sciences, Office of Science, U.S. Department of Energy through award No. DE-
FG02-04ER15571, is gratefully acknowledged. Support for K.B. was provided by DOE. 

\bibliography{MFCS2}

\end{document}